\documentclass[prd,aps,twocolumn,showpacs,preprintnumbers,amsmath,amssymb]{revtex4-2}
\usepackage[utf8]{inputenc}
\usepackage{graphicx}
\usepackage[breaklinks,colorlinks=true,linkcolor=blue,citecolor=blue,urlcolor=blue]{hyperref}
\usepackage{bm}
\usepackage{physics}
\usepackage{siunitx}
\usepackage{mathtools}
\usepackage{amsmath}

\newcommand{\Lag}{\mathcal{L}}
\graphicspath{{./}{figs/}{paper_eta/figs/}}
\newcommand{\ord}[1]{\mathcal{O}\!\left(#1\right)}
\newcommand{\Slash}[1]{\ooalign{\hfil/\hfil\crcr$#1$}}

\begin{document}

\title{Chiral anomaly and nucleon spin decomposition in the chiral quark soliton model}

\author{Josuke Minamiguchi}
\email{minamiguchi@nt.phys.s.u-tokyo.ac.jp}
\affiliation{Department of Physics, The University of Tokyo, 
  7-3-1 Hongo, Bunkyo-ku, Tokyo 113-0033, Japan}

\author{Tomoya Uji}
\email{uji@nt.phys.s.u-tokyo.ac.jp}
\affiliation{Department of Physics, The University of Tokyo, 
  7-3-1 Hongo, Bunkyo-ku, Tokyo 113-0033, Japan}

\begin{abstract}
We investigate how the chiral anomaly influences the decomposition of nucleon spin between quark spin and orbital motion.  We extend the chiral quark soliton model by including a flavor-singlet pseudoscalar field whose mass is generated by the chiral anomaly.  Within this framework, we calculate the quark and antiquark helicity and orbital angular momentum distributions, accounting for the singlet field induced by the rotation of the soliton.  We vary the anomaly-induced mass while keeping the remaining model parameters fixed to examine how the anomaly affects the spin decomposition.  The induced field enhances the quark helicity contribution relative to the conventional pion-only model.  Increasing the mass suppresses this enhancement and increases the orbital contribution, while the two contributions continue to account for the nucleon spin.  This redistribution remains modest over the parameter values considered and does not qualitatively alter the spin decomposition of the conventional model.
\end{abstract}

\maketitle

\section{Introduction}
\label{sec:introduction}

The Electron-Ion Collider will provide new opportunities to explore the spin structure of the nucleon through polarized inclusive and exclusive scattering~\cite{Accardi:2012qut, AbdulKhalek:2021gbh}.  A central goal of this program is to determine how the nucleon spin is shared among the helicity and orbital angular momentum (OAM) of quarks and gluons~\cite{Jaffe:1989jz, Ji:1996ek, Leader:2013jra}; see Ref.~\cite{Hatta:2026txn} for the state-of-the-art overview.  Alongside the experimental determination of these contributions, understanding the nonperturbative mechanisms that govern their relative magnitudes remains an important theoretical task.  The flavor-singlet sector is particularly interesting in this respect because the quark helicity is connected to gluon topology through the chiral anomaly~\cite{Jaffe:1989jz, Shore:2007yn, Zahed:2026wag}.

The role of the $U(1)_A$ anomaly in the proton spin problem has been investigated since the early studies of polarized deep-inelastic scattering (DIS)~\cite{Kodaira:1979pa, Altarelli:1988nr, Carlitz:1988ab}.  The first moment of the flavor-singlet quark helicity distribution is given by the forward nucleon matrix element of the singlet axial current.  Topological charge screening in the QCD vacuum was subsequently proposed as a nonperturbative explanation for the suppression of the singlet axial charge~\cite{Shore:1990zu, Shore:1991dv, Narison:1994hv}.  This explanation has, however, been debated in QCD sum-rule analyses~\cite{Ioffe:1998rh, Narison:1998aq} and questioned by a nonlocal chiral quark model calculation~\cite{Dorokhov:2003kf}.

More recently, the connection between the chiral anomaly and nucleon spin structure has been explored in polarized DIS, including the small-$x$ regime~\cite{Tarasov:2020cwl, Tarasov:2021yll, Tarasov:2025mvn}, in deeply virtual Compton scattering and generalized parton distributions~\cite{Hatta:2020ltd, Bhattacharya:2022xxw, Bhattacharya:2023wvy, Bhattacharya:2024geo}, and through the topological dipole structure of polarized nucleons~\cite{Fukushima:2026sot}.  These developments highlight the role of flavor-singlet pseudoscalar dynamics and motivate a quantitative investigation of how it affects the size of the quark helicity in the nucleon.  A framework with explicit quark degrees of freedom is particularly useful for addressing this question.

The chiral quark soliton model (CQSM) provides such a framework~\cite{Diakonov:1987ty, Diakonov:1997sj}.  Based on spontaneous chiral symmetry breaking and the large-$N_c$ limit, it describes the nucleon in terms of constituent quarks in a self-consistent chiral mean field, including both the discrete valence level and the distorted Dirac sea.  Its field-theoretical formulation allows both quark and antiquark distributions to be calculated consistently at a low model scale.  In particular, the quark helicity and OAM distributions can be calculated within the same framework.  In the conventional CQSM without explicit gluons, the nucleon spin is entirely carried by the spin and OAM of quarks and antiquarks~\cite{Wakamatsu:1990ud, Wakamatsu:1999nj}.  Recent studies have further explored the role of instanton-induced chiral interactions in quark spin--orbit correlations~\cite{Kim:2024cbq} and flavor-nonsinglet OAM~\cite{Kim:2026gat}.  The instanton-vacuum approach also provides a description of the singlet axial charge consistent with the $U(1)_A$ anomaly at leading order in $1/N_c$~\cite{Diakonov:1995qy}.  The standard two-flavor formulation, however, retains only the pion field and does not explicitly resolve the response of the flavor-singlet pseudoscalar mode.

In this work, we extend the two-flavor CQSM by explicitly retaining the flavor-singlet pseudoscalar field, $\eta_0$.  This field represents the would-be Nambu--Goldstone mode of singlet axial symmetry, which acquires a mass through the $U(1)_A$ anomaly~\cite{Witten:1979vv, Veneziano:1979ec}.  The anomalous axial Ward identity links this pseudoscalar sector to the singlet axial charge, and hence to the quark helicity contribution to the spin~\cite{Shore:1990zu, Shore:1991dv}.  Retaining this mode with explicit quark degrees of freedom allows us to investigate how its response in a polarized nucleon affects both quark helicity and OAM.  We vary its anomaly-induced mass, $m_0$, while keeping the remaining model parameters fixed to probe how anomalous $U(1)_A$ breaking influences the partition of the nucleon spin between these two contributions.  We find that increasing $m_0$ reduces the quark helicity contribution over the parameter values considered.  The resulting reduction in the quark helicity contribution is, however, modest.

\section{Theoretical Framework}\label{sec:theoretical}
In this section, we extend the CQSM to include the flavor-singlet pseudoscalar field, $\eta_0$, and its anomaly-induced mass.  We formulate the collective rotational dynamics of the extended model and derive expressions for the quark helicity and orbital angular momentum distributions, including the contributions from the rotation-induced singlet field.

\subsection{The chiral quark soliton model with \texorpdfstring{$\eta_0$}{eta0} meson}

The CQSM is an effective model of baryons based on spontaneous chiral symmetry breaking and the large-$N_c$ limit of QCD~\cite{Diakonov:1997sj}.  In the standard two-flavor formulation, constituent quarks with a dynamically generated mass interact with a pion field, $U\in SU(2)$.  At leading order in the $1/N_c$ expansion, the baryon is described by $N_c$ valence quarks occupying a discrete bound-state level in a self-consistent pion mean field, together with the Dirac sea.  Nucleon states with definite spin and isospin are obtained by quantizing the collective rotations of the soliton.

Although the standard two-flavor CQSM retains only the pion field, a flavor-singlet pseudoscalar degree of freedom is already present in the bosonized instanton-vacuum effective theory underlying the model~\cite{Diakonov:1997sj}.  In the usual low-momentum reduction, this mode is frozen out because its anomaly-induced mass is of order the intrinsic ultraviolet cutoff, set by the inverse average instanton size.  Here, we retain the singlet field explicitly to study how its anomaly-induced dynamics affects the spin structure of the nucleon.  We therefore work with the $U(2)$ chiral field
\begin{equation}
    \mathcal{U}(x) = e^{i\eta_0(x)/f_0} U(x)\,,
\end{equation}
where $f_0$ denotes the singlet decay constant.  Relating $\eta_0$ to the physical $\eta$ and $\eta'$ mesons requires a three-flavor treatment incorporating $\eta$--$\eta'$ mixing.  Such an extension is beyond the scope of the present work.  

The flavor-singlet sector must also account for the anomalous breaking of $U(1)_A$ symmetry in QCD.  To incorporate this effect consistently with the anomalous singlet axial Ward identity, we supplement the effective action with the leading Witten--Di Vecchia--Veneziano potential~\cite{Witten:1979vv, Veneziano:1979ec, DiVecchia:1980yfw, Witten:1980sp}.  We set the QCD vacuum angle to $\theta = 0$ throughout this work.  On the branch containing $\eta_0 = 0$, the anomaly term takes the form
\begin{equation}
    \Lag_A = -\frac{\chi_{\mathrm{YM}}}{2}\left(-i\ln\det\mathcal{U}\right)^2 = -\frac{1}{2}m_0^2\eta_0^2\,,
    \label{eq:anomaly-potential}
\end{equation}
where $\chi_{\mathrm{YM}}$ is the topological susceptibility of pure Yang--Mills theory and
\begin{equation}
    m_0^2 = \frac{4\chi_{\mathrm{YM}}}{f_0^2}\,.
    \label{eq:m0-chi}
\end{equation}
This term generates a nonzero singlet mass even in the chiral limit.  
The quantity $\chi_{\mathrm{YM}}$ should be distinguished from the topological susceptibility of full QCD, which vanishes in this limit.  
In the present study, we treat $m_0$ as a tunable model parameter controlling the strength of anomaly-induced $U(1)_A$ breaking, rather than fixing it to a physical pseudoscalar-meson mass.

With the anomaly term included, the effective Minkowski Lagrangian for constituent quarks coupled to the $U(2)$ chiral field reads
\begin{align}
    \Lag &= \bar\psi\left(i\Slash{\partial} - M\mathcal{U}^{\gamma_5}\right)\psi - \frac{1}{2}m_0^2\eta_0^2\,,
  \label{eq:cqsm-lagrangian}
\end{align}
with
\begin{equation}
    \mathcal{U}^{\gamma_5} = \exp\left(i\gamma_5\frac{\eta_0}{f_0}\right)U^{\gamma_5}\,, \qquad U^{\gamma_5} = \exp\left(i\gamma_5\frac{\tau^a\pi^a}{f_\pi}\right)\,.
  \label{eq:chiral-field}
\end{equation}
Here, $M$ is the constituent quark mass, and $f_\pi$ is the pion decay constant.  The kinetic terms for pseudoscalar mesons arise from the derivative expansion of the quark determinant~\cite{Diakonov:1997sj}.  After the quarks are integrated out, the effective action becomes
\begin{equation}
    S_{\mathrm{eff}}[U, \eta_0] = - iN_c\Tr\ln\left(i\Slash{\partial} - M\mathcal{U}^{\gamma_5}\right) - \frac{m_0^2}{2}\int d^4x\,\eta_0^2\,.
    \label{eq:effective-action}
\end{equation}
The trace is over space-time, Dirac, and flavor indices, while the color trace gives the explicit factor $N_c$.  The extension in Eq.~\eqref{eq:cqsm-lagrangian} treats $\eta_0$ as a massive response field, not as an additional collective zero mode; the pion-only CQSM is recovered continuously for $m_0\to\infty$.


\subsection{Static hedgehog and single-particle spectrum}

At large $N_c$, we evaluate the path integral over the chiral field at its saddle point~\cite{Diakonov:1987ty, Christov:1995vm}.  The classical pion field is the hedgehog
\begin{equation}
    U_c^{\gamma_5}(\bm{r}) = \exp\left[i\gamma_5\bm\tau\cdot\hat{\bm{r}}F(r)\right]\,,
  \label{eq:hedgehog}
\end{equation}
and parity forbids a static $\eta_0$ background, so that $\eta_{0, c} = 0$, where the subscript $c$ labels the static saddle-point configuration.  The quark single-particle Dirac Hamiltonian in this background is
\begin{align}
    H_c &= - i\gamma^0\gamma^k\partial_k + \gamma^0MU_c^{\gamma_5}\,,\qquad H_c\ket{n} = E_n\ket{n}\,.
  \label{eq:dirac-spectrum}
\end{align}
The hedgehog is invariant under simultaneous rotations in space and in isospin, so $H_c$ commutes with the grand spin, $\bm{K} = \bm{L} + \bm{S} + \bm{T}$, the sum of the orbital angular momentum, $\bm{L}$, the quark spin, $\bm{S}$, and the isospin, $\bm{T}$.  It also commutes with the parity, so each state carries a grand spin, $\bm{K}$, and a parity, $\Pi$.  

We determine the profile, $F(r)$, self-consistently by requiring stationarity of the classical soliton mass,
\begin{equation}
    M_{\mathrm{cl}} = N_c\sum_{n\in\mathrm{occ}}E_n - N_c\sum_{n\in\mathrm{occ}}E_n\Big|_{F\equiv0}\,,
    \label{eq:Mcl}
\end{equation}
where ``occ'' denotes the quark single-particle states occupied in the baryon, namely the discrete valence level and the negative-energy continuum.  The second term subtracts the vacuum contribution at $F \equiv 0$, and the spectral sums are regularized as described in Sec.~\ref{sec:reg}.

\subsection{Collective rotation and the induced singlet field}

To construct nucleon states with definite spin and isospin from the static hedgehog, we quantize the rotational zero modes of the soliton~\cite{Diakonov:1987ty, Wakamatsu:1990ud}.  These collective rotations are parametrized by a slowly varying matrix $A(t)\in SU(2)$,
\begin{equation}
    U^{\gamma_5}(\bm{r}, t) = A(t)U_c^{\gamma_5}(\bm{r})A^\dagger(t)\,,\qquad A^\dagger\dot{A} = \frac{i}{2}\Omega^a\tau^a\,,
  \label{eq:collective-rotation}
\end{equation}
where $\Omega^a$ are the components of the angular velocity in the body-fixed frame.

Although the singlet field vanishes in the static hedgehog, collective rotation can induce a nonzero response.  At linear order in the angular velocity, parity and the hedgehog symmetry constrain its angular dependence in the body-fixed frame, leading to the parametrization
\begin{equation}
    \eta_0(\bm{r}, t) = \eta(r)\Omega^a(t)\hat{r}^a + \ord{\Omega^2}\,.
\label{eq:eta-profile}
\end{equation}
An analogous rotation-induced $\eta_0$ profile occurs in the Skyrme model with explicit flavor singlet mesons~\cite{Jain:1987sz, Meissner:1988iv, Schechter:1999hg, Fukushima:2026sot}.  In the rotating frame, the Dirac operator up to second order in the angular velocity reads
\begin{align}
    \gamma^0\mathcal{D} &= D_0 - V_\Omega - V_\eta - V_{\eta^2} + \ord{\Omega^3}\,,
    \label{eq:rotating-dirac}
\end{align}
where
\begin{align}
    D_0 &= i\partial_t - H_c\,,\qquad V_\Omega = \frac{1}{2}\Omega^a\tau^a\,, \\
    V_\eta &= i\gamma^0\gamma_5\frac{M\eta_0}{f_0}U_c^{\gamma_5}\,,\qquad V_{\eta^2} = -\frac{M\eta_0^2}{2f_0^2}\gamma^0U_c^{\gamma_5}\,.
  \label{eq:collective-vertices}
\end{align}

\subsection{Moment of inertia}

We now determine the induced singlet profile and its contribution to the moment of inertia by expanding the effective action in Eq.~\eqref{eq:effective-action} through second order in the collective angular velocity $\Omega$.  The first-order contribution is
\begin{equation}
    S_{\mathrm{eff}}^{(1)} = iN_c\Tr\left[D_0^{-1}(V_\Omega + V_\eta)\right]
    = 0\,,
    \label{eq:collective-action-linear}
\end{equation}
where the single-insertion terms vanish by the symmetries of the static hedgehog.

The quadratic contribution can be organized as
\begin{equation}
    S_{\mathrm{eff}}^{(2)} = S_{\Omega\Omega} + S_{\Omega\eta} + S_{\eta\eta}\,,
    \label{eq:collective-action-quadratic}
\end{equation}
with
\begin{align}
    S_{\Omega\Omega} &= \frac{iN_c}{2}\Tr\left(D_0^{-1}V_\Omega D_0^{-1}V_\Omega\right)\,,
       \label{eq:action-omega-omega} \\
    S_{\Omega\eta} &= iN_c\Tr\left(D_0^{-1}V_\Omega D_0^{-1}V_\eta\right)\,,
       \label{eq:action-omega-eta} \\
    S_{\eta\eta} &= \frac{iN_c}{2}\Tr\left(D_0^{-1}V_\eta D_0^{-1}V_\eta\right) + iN_c\Tr\left(D_0^{-1}V_{\eta^2}\right) \notag \\
    &\qquad - \frac{1}{2}m_0^2\int d^4x\,\eta_0^2\,.
       \label{eq:action-eta-eta}
\end{align}

To evaluate the functional traces, we use the spectral representation of the static propagator,
\begin{equation}
\begin{split}
    &\mel{n}{D_0^{-1}(\omega)}{m} = \delta_{nm}g_n(\omega)\,, \\
    &g_n(\omega) = \frac{1}{\omega - E_n  + i0^+(1 - 2\nu_n)}\,,
    \label{eq:static-propagator-spectral}
\end{split}
\end{equation}
where $\omega$ is the frequency variable conjugate to time, $t$, and $0^+$ denotes a positive infinitesimal specifying the pole prescription.  The occupation number is $\nu_n = 1$ for occupied states
(including the valence level) and $\nu_n = 0$ for unoccupied states.  At leading adiabatic order, the response functions are evaluated at zero external frequency.  The internal frequency integrals then give
\begin{align}
    i\int\frac{d\omega}{2\pi}\,g_n(\omega)g_m(\omega) &= \frac{\nu_n - \nu_m}{E_m - E_n}\,,
    \label{eq:frequency-integral-two} \\
    i\int\frac{d\omega}{2\pi}\,e^{i\omega 0^+}g_n(\omega) &= - \nu_n\,.
    \label{eq:frequency-integral-one}
\end{align}
We first display the resulting unregularized spectral expressions.  The Dirac-sea contributions are to be regularized consistently with the effective action, as described in Sec.~\ref{sec:reg}.

Performing the frequency integrals and using the Hermiticity of the vertices, we obtain
\begin{align}
    S_{\Omega\Omega} &= N_c\int dt\sum_{n\in\mathrm{occ}, \,m\in\mathrm{non}}\frac{\left|\mel{m}{V_\Omega}{n}\right|^2}{E_m-E_n}\,,
       \label{eq:action-omega-omega-spectral} \\
    S_{\Omega\eta} &= 2N_c\int dt\sum_{n\in\mathrm{occ},\,m\in\mathrm{non}}\Re\frac{\mel{n}{V_\Omega}{m}\mel{m}{V_\eta}{n}}{E_m - E_n}\,,
       \label{eq:action-omega-eta-spectral} \\
    S_{\eta\eta} &= N_c\int dt\sum_{n\in\mathrm{occ}, \,m\in\mathrm{non}}\frac{\left|\mel{m}{V_\eta}{n}\right|^2}{E_m-E_n} \notag \\
    &\qquad - N_c\int dt\sum_{n\in\mathrm{occ}}\mel{n}{V_{\eta^2}}{n} - \frac{1}{2}m_0^2\int d^4x\,\eta_0^2\,,
       \label{eq:action-eta-eta-spectral}
\end{align}
where ``occ'' denotes the occupied quark single-particle states ($\nu_n=1$), comprising the discrete valence level and the negative-energy continuum, and ``non'' denotes the complementary set of unoccupied states ($\nu_n=0$).  For the two-insertion terms, contributions with equal occupations vanish owing to the factor $\nu_n-\nu_m$ from the frequency integral.

Using Eq.~\eqref{eq:eta-profile}, we obtain the matrix elements of the singlet vertex,
\begin{equation}
    \mel{m}{V_\eta}{n} = \Omega^b\int_0^\infty dr\,r^2\eta(r)y_{mn}^b(r)\,,
    \label{eq:eta-vertex-matrix-element}
\end{equation}
with
\begin{align}
    &y_{mn}^a(r) = \int d\Omega_{\bm{r}}\,\Phi_m^\dagger(\bm{r})\hat{r}^a\Gamma_\eta(\bm{r})\Phi_n(\bm{r})\,,  \\
    &\Gamma_\eta = i\gamma^0\gamma_5\frac{M}{f_0}U_c^{\gamma_5}\,.
  \label{eq:eta-vertex}
\end{align}
Here, $\Phi_n(\bm{r}) = \braket{\bm{r}}{n}$ is the single-quark wave function.
Substituting this expression into
Eq.~\eqref{eq:action-omega-eta-spectral} gives
\begin{align}
    S_{\Omega\eta} &= \int dt\,\Omega^2
       \int_0^\infty dr\,r^2\eta(r)j_\eta(r)\,,
       \label{eq:action-omega-eta-radial}
\end{align}
where 
\begin{equation}
    j_\eta(r) = \frac{N_c}{3}\sum_{nm}'\Re\frac{\mel{n}{\tau^a}{m}y_{mn}^a(r)}{E_m - E_n}\,,
    \label{eq:eta-source}
\end{equation}
with $\sum_{nm}' = \sum_{n\in\mathrm{occ}, m\in\mathrm{non}}$.

Similarly, the singlet-quadratic contribution becomes
\begin{align}
    S_{\eta\eta} &= \int dt\,\Omega^2\Bigg[\int_0^\infty dr\,r^2\int_0^\infty dr'\,{r'}^2\eta(r)q_0(r,r')\eta(r') \notag \\
    &\qquad +\int_0^\infty dr\,r^2q_1(r)\eta^2(r) - \frac{2\pi}{3}m_0^2\int_0^\infty dr\,r^2\eta^2(r)\Bigg]\,,
    \label{eq:action-eta-eta-radial}
\end{align}
with
\begin{align}
    &q_0(r, r') = \frac{N_c}{3}\sum_{nm}'\Re\frac{y_{nm}^a(r)y_{mn}^a(r')}{E_m - E_n}\,,
    \label{eq:q0-kernel} \\
    &q_1(r) = - \frac{N_c}{3}\sum_{n\in\mathrm{occ}}\int d\Omega_{\bm{r}}\,\Phi_n^\dagger(\bm{r})\Gamma_{\eta^2}(\bm{r})\Phi_n(\bm{r})\,,
    \label{eq:q1-kernel} \\
    &\Gamma_{\eta^2} = - \gamma^0\frac{M}{2f_0^2}U_c^{\gamma_5}\,.
\end{align}
Here $q_0$ arises from two insertions of $V_\eta$, whereas $q_1$ arises from one insertion of $V_{\eta^2}$.

Introducing the radial inner product $\expval{f, g} = \int_0^\infty dr\,r^2f(r)g(r)$, we write
\begin{equation}
    S_{\eta\eta} = - \frac{1}{2}\int dt\,\Omega^2\expval{\eta,\mathcal K\eta}\,,
\end{equation}
where
\begin{equation}
    \mathcal K(r, r') = \left[\frac{4\pi}{3}m_0^2 - 2q_1(r)\right]\frac{\delta(r - r')}{r^2} - 2q_0(r, r')\,.
    \label{eq:eta-kernel}
\end{equation}

The ordinary rotational inertia~\cite{Wakamatsu:1990ud}
is
\begin{align}
  I_{\Omega\Omega} &= \frac{N_c}{6}\sum_{nm}'\frac{\mel{n}{\tau^a}{m}\mel{m}{\tau^a}{n}}{E_m - E_n}\,.
  \label{eq:ordinary-inertia}
\end{align}

Collecting these contributions and adding the static term $S_{\mathrm{eff}}^{(0)} = - \int dt\,M_{\mathrm{cl}}$, we obtain the collective Lagrangian
\begin{equation}
    L_{\mathrm{coll}} = - M_{\mathrm{cl}} + \frac{\Omega^2}{2}\left[I_{\Omega\Omega} + 2\expval{\eta, j_\eta} - \expval{\eta, \mathcal K\eta}\right]\,.
    \label{eq:collective-lagrangian}
\end{equation}

The profile $\eta(r)$ is determined by stationarity of the quadratic action,
\begin{equation}
  \int_0^\infty dr'\,{r'}^2\mathcal{K}(r, r')\eta(r') = j_\eta(r)\,.
  \label{eq:eta-response-equation}
\end{equation}
On this solution, the total moment of inertia is
\begin{equation}
    I = I_{\Omega\Omega} + I_\eta\,,
    \label{eq:total-inertia}
\end{equation}
with
\begin{equation}
  I_\eta = \langle\eta, j_\eta\rangle\,.
\end{equation}
Quantizing the collective rotation gives the nucleon mass
\begin{equation}
  M_N = M_{\mathrm{cl}} + \frac{J(J+1)}{2I} = M_{\mathrm{cl}} + \frac{3}{8I}\,,
  \label{eq:MN}
\end{equation}
with $J = \frac{1}{2}$.


\subsection{Regularization}
\label{sec:reg}

The quark loop in Eq.~\eqref{eq:effective-action} is ultraviolet divergent.  Let $\Gamma[U, \eta_0; \mu] = - iN_c\Tr\ln(i\Slash{\partial} - \mu\,\mathcal{U}^{\gamma_5})$ denote that loop with the constituent quark mass replaced by $\mu$.  We regularize it with the double Pauli--Villars subtraction~\cite{Kubota:1999hx},
\begin{equation}
    \Gamma^{\mathrm{reg}}[U, \eta_0] = \Gamma[U, \eta_0; M] - \sum_{i=1}^{2}c_i\,\Gamma[U, \eta_0; \Lambda_i]\,.
  \label{eq:pvaction}
\end{equation}
The Pauli--Villars subtraction is applied only to the quark determinant.  The anomaly-induced mass term is retained separately, so that the regularized effective action reads
\begin{equation}
    S_{\mathrm{eff}}^{\mathrm{reg}}[U, \eta_0] = \Gamma^{\mathrm{reg}}[U, \eta_0] - \frac{1}{2}m_0^2\int d^4x\,\eta_0^2\,.
\end{equation}  
The substitution $M\to\mu$ acts on $H_c$ and on the $\eta$ vertices of Eq.~\eqref{eq:collective-vertices}.  The cutoff dependence cancels if $\sum_i c_i\Lambda_i^2 = M^2$ and $\sum_i c_i\Lambda_i^4 = M^4$, and these two conditions fix $c_1$ and $c_2$ in terms of $\Lambda_1$ and $\Lambda_2$.   The masses $\Lambda_1$ and $\Lambda_2$ are fixed by the physical values of the pion decay constant and the chiral condensate~\cite{Kubota:1999hx}.  The constituent mass $M$ is then the only free parameter.  Alternative subtraction schemes are discussed in Refs.~\cite{Weigel:1999pc, Takyi:2019ahv}.

In the vacuum, $F\equiv0$, the source $j_\eta$ of Eq.~\eqref{eq:eta-source} vanishes.  Hence $\eta(r)$ vanishes, and the $\eta$ vertex has no vacuum counterpart.   We write $\mathcal{Q}[\mu; F]$ for any quantity $\mathcal{Q}$ computed with quark mass $\mu$ and profile $F$.  The vacuum-subtracted, Pauli--Villars-regularized form is
\begin{equation}
  \begin{aligned}
    \mathcal{Q}^{\mathrm{reg}} &= \mathcal{Q}[M; F]-\mathcal{Q}[M; 0] \\
    &\qquad - \sum_{i = 1}^{2}c_i\left\{\mathcal{Q}[\Lambda_i; F] - \mathcal{Q}[\Lambda_i; 0]\right\}\,.
  \end{aligned}
  \label{eq:Qreg}
\end{equation}
The spectrum is generated three times, at $\mu\in\{M, \Lambda_1, \Lambda_2\}$ with the same profile.  The valence level is not regularized~\cite{Ossmann:2004bp}.  Equation~\eqref{eq:eta-response-equation} is solved once, and the same $\eta(r)$ is used at all three masses.

\subsection{Quark helicity PDF}
\label{sec:pol}

The quark helicity PDF is defined by the light-cone correlator of the axial-vector bilinear.  In the nucleon rest frame, the isoscalar quark helicity PDF is~\cite{Wakamatsu:1998rx}
\begin{widetext}
\begin{equation}
    \left[\Delta u + \Delta d\right]\!(x) = \frac{1}{4\pi}\int dz^{0}\,e^{ixM_{N}z^{0}}\,\mel{\bm P = \bm 0, S_3}{\psi^{\dagger}(0)\left(1 + \gamma^{0}\gamma^{3}\right)\gamma_5\,\psi(z)}{\bm P = \bm0, S_3}\Big|_{z^{3} = - z^{0}, \,z_\perp = 0}\,,
    \label{eq:defpol}
\end{equation}
\end{widetext}
where $\ket{\bm P, S_3}$ denotes a nucleon state with three-momentum $\bm P$ and spin $S_3$.  The support is extended to $-1\le x\le1$, where the negative-$x$ region carries the antiquark distribution~\cite{Diakonov:1997vc, Wakamatsu:1998rx},
\begin{equation}
    \left[\Delta\bar{u} + \Delta\bar{d}\right]\!(x) = \left[\Delta u + \Delta d\right]\!(-x)\,, \qquad 0 < x < 1\,,
    \label{eq:antiqpol}
\end{equation}
and the first moment is the flavor singlet axial charge,
\begin{equation}
    \Delta\Sigma = \int_{-1}^{1}dx\,\left[\Delta u+\Delta d\right]\!(x) = g_A^{(0)}\,.
    \label{eq:DeltaSigma}
\end{equation}
In the parton-distribution formulas below, we replace $M_N$ by the classical soliton mass $M_{\mathrm{cl}}$. 
The rotational mass correction in Eq.~\eqref{eq:MN} is $\ord{1/N_c}$ and thus represents a relative correction of $\ord{1/N_c^2}$ to $M_{\mathrm{cl}}=\ord{N_c}$.  Including this correction in the kinematic mass factors and Fourier phase would affect the distributions only beyond the accuracy of the present calculation, which retains terms linear in the collective angular velocity.

To evaluate $[\Delta u + \Delta d](x)$ in this model, we introduce two single-particle operators: $\mathrm{O}^S_n(x)$ for the helicity insertion and $Y^a$ for the coupling to the induced singlet field.  The former is defined as~\cite{Diakonov:1997vc, Wakamatsu:1998rx}
\begin{equation}
    \mathrm{O}^S_n(x) = \left(1+\gamma^0\gamma^3\right)\gamma_5\,\delta_n\,,
    \label{eq:OS-n}
\end{equation}
with
\begin{equation}
    \delta_n = \delta\!\left(xM_{\mathrm{cl}} - E_n - p^3\right)\,,
    \label{eq:delta-n}
\end{equation}
where $p^3$ is the third component of the quark momentum operator.

For the latter, we use Eq.~\eqref{eq:eta-profile} to factor the angular velocity out of the singlet vertex in Eq.~\eqref{eq:collective-vertices},
\begin{equation}
    V_\eta = \Omega^aY^a\,, \qquad Y^a = \eta(r)\hat r^a\Gamma_\eta\,.
    \label{eq:Y-operator}
\end{equation}
Its matrix elements in the static single-quark basis are
\begin{equation}
    \mel{m}{Y^a}{n} = \int_0^\infty dr\,r^2\eta(r)y_{mn}^a(r)\,,
\end{equation}
with $\Gamma_\eta$ and $y_{mn}^a(r)$ defined above.

We now evaluate $[\Delta u + \Delta d](x)$ to first order in the collective angular velocity.  At this order, contributions arise from the expansion of both the quark propagator in the rotating frame and the collective rotation matrices in the bilocal operator.

The propagator expansion involves single insertions of $V_\Omega = \Omega^a\tau^a/2$ and $V_\eta = \Omega^aY^a$.  The former gives the conventional $\{A, B\}$ contribution in the notation of Ref.~\cite{Wakamatsu:1998rx}, while the latter gives the additional contribution from the induced singlet field.  The bilocal operator in Eq.~\eqref{eq:defpol}, on the other hand, contains the product $A^\dagger(0)A(z^0)$ of rotation matrices evaluated at different times. Expanding this product to first order in $\Omega$ gives the $C$ contribution~\cite{Pobylitsa:1998tk}.  Collecting these three contributions, we obtain
\begin{widetext}
\begin{align}
    \left[\Delta u + \Delta d\right]\!(x) &= \left[\Delta u + \Delta d\right]^{\{A, B\}'}_\Omega\!(x) + \left[\Delta u + \Delta d\right]^{C'}_\Omega\!(x) + \left[\Delta u + \Delta d\right]_\eta\!(x) + \ord{\Omega^2}\,,
\end{align}
where
\begin{align}
    \left[\Delta u + \Delta d\right]^{\{A, B\}'}_\Omega\!(x) &= M_{\mathrm{cl}}\frac{N_c}{2I}\sum_{\substack{m = \mathrm{all},\,n\in\mathrm{occ} \\ 
    (E_m \neq E_n)}}\frac{1}{E_n - E_m}\mel{n}{\tau^3}{m}\mel{m}{\mathrm{O}^S_n\!(x)}{n}\,,
    \label{eq:pol-AB} \\
    \left[\Delta u + \Delta d\right]^{C'}_\Omega\!(x) &= \frac{d}{dx}\frac{N_c}{4I}\!\sum_{\substack{m = \mathrm{all},\,n\in\mathrm{occ} \\ 
    (E_m \neq E_n)}}\!\mel{n}{\tau^3}{m}\mel{m}{\mathrm{O}^S_n\!(x)}{n}\,,
    \label{eq:pol-C} \\
    \left[\Delta u + \Delta d\right]_\eta\!(x) &= M_{\mathrm{cl}}\frac{N_c}{I}\!\sum_{\substack{m = \mathrm{all},\,n\in\mathrm{occ} \\ 
    (E_m \neq E_n)}}\!\frac{\mel{n}{Y^3}{m}\mel{m}{\mathrm{O}^S_n\!(x)}{n}}{E_n - E_m} - \frac{d}{dx}\frac{N_c}{2I}\!\sum_{\substack{m = \mathrm{all},\,n\in\mathrm{occ} \\
    (E_m = E_n)}}\!\mel{n}{Y^3}{m}\mel{m}{\mathrm{O}^S_n\!(x)}{n}\,.
    \label{eq:pol-eta} 
\end{align}
\end{widetext}
The primes on the conventional $\{A, B\}$ and $C$ contributions indicate that the terms with $E_m=E_n$ have already been canceled between them.  The remaining sums in these two contributions therefore run only over nondegenerate pairs.  

In contrast, the singlet contribution at first order in $\Omega$ arises solely from a single insertion of $V_\eta$ in the quark propagator.  Since $V_\eta=\ord{\Omega}$, combining this insertion with the first-order expansion of the endpoint rotation matrices would give an $\ord{\Omega^2}$ contribution, beyond the accuracy retained here.  Thus, there is no corresponding $C$ term at $\ord{\Omega}$, and the degenerate contribution from $V_\eta$ must be retained.  This contribution is evaluated using the coincident-energy limit
\begin{equation}
    \lim_{E_m\to E_n}\frac{\delta_n-\delta_m}{E_n-E_m} = -\frac{1}{M_{\mathrm{cl}}}\frac{d}{dx}\delta_n\,,
    \label{eq:lhopital}
\end{equation}
which yields the derivative term in Eq.~\eqref{eq:pol-eta}.

In Eqs.~\eqref{eq:pol-AB}--\eqref{eq:pol-eta}, the index $n$ runs over the single-quark states occupied in the baryon.  The intermediate-state index $m$, by contrast, runs over both occupied and unoccupied states, subject to the energy conditions specified in each sum.
We refer to this form of the expressions as the occupied-state representation.

The same helicity distribution can alternatively be expressed through sums over the complementary set of single-quark states that are not occupied in the baryon, denoted by $\mathrm{non}$.  This alternative form, which we call the unoccupied-state representation, is obtained by making the replacement
\begin{equation}
    \sum_{n\in\mathrm{occ}}
    \;\longrightarrow\;
    -\sum_{n\in\mathrm{non}}\,,
    \label{eq:occ-non-replacement}
\end{equation}
while the sum over $m$ and the energy conditions are left unchanged.
The two representations are alternative expressions for the same distribution.


\subsection{Quark orbital angular momentum}
\label{sec:OAM}

For the quark orbital angular momentum, we adopt the definition of Ref.~\cite{Wakamatsu:1999nj}, formulated in the light-cone gauge $A^+=0$,
\begin{widetext}
\begin{equation}
    q_L(x) = \frac{1}{\sqrt{2}\,p^{+}}\int_{-\infty}^{\infty}\frac{d\lambda}{2\pi}\,e^{i\lambda x}\,\mel{\bm P = \bm0, S_3}{\psi_{+}^{\dagger}(0)\,i\!\left(X^{1}\partial^{2}-X^{2}\partial^{1}\right)\psi_{+}(\lambda n)}{\bm P=\bm0, S_3}\,,
    \label{eq:defqL}
\end{equation}
\end{widetext}
where $\psi_\pm = P_\pm\psi$ with $P_\pm=\tfrac12\gamma^{\mp}\gamma^{\pm}$, and $\bm{X} = r\hat{\bm{r}}$ is the position operator.  As in Eq.~\eqref{eq:antiqpol} the support is extended to $-1\le x\le1$,
\begin{equation}
    \bar{q}_L(x) = q_L(-x)\,, \qquad 0 < x < 1\,,
    \label{eq:antiqL}
\end{equation}
and the first moment is the quark orbital angular momentum,
\begin{equation}
    L_q = \int_{-1}^{1}dx\,q_L(x)\,.
    \label{eq:Lq}
\end{equation}
In the model, Eq.~\eqref{eq:defqL} is evaluated in the same way as Eq.~\eqref{eq:defpol}, with $\mathrm{O}^S_n\!(x)$ replaced by
\begin{align}
    \mathrm{O}^L_n\!(x) = \left(1 + \gamma^0\gamma^3\right)
    i\!\left(X^{1}\partial^{2}-X^{2}\partial^{1}\right)\delta_n\,.
    \label{eq:OL-n}
\end{align}
Like $[\Delta u + \Delta d](x)$, $q_L(x)$ starts at order $\Omega$ and receives both an $\{A,B\}$ and a $C$ term, with the same prefactors as in Eqs.~\eqref{eq:pol-AB} and \eqref{eq:pol-C},
\begin{widetext}
\begin{align}
    q_L\!(x) &= q_{L, \Omega}^{\{A, B\}'}\!(x) +  q_{L,\Omega}^{C'}(x) + q_{L,\eta}(x) + \ord{\Omega^2}\,,
\end{align}
where
\begin{align}
    q_{L,\Omega}^{\{A, B\}'}\!(x) &= M_{\mathrm{cl}}\frac{N_c}{2I}\sum_{\substack{m = \mathrm{all},\,n\in\mathrm{occ} \\ 
    (E_m \neq E_n)}}\frac{1}{E_n - E_m}\,\mel{n}{\tau^3}{m}\mel{m}{\mathrm{O}^L_n\!(x)}{n}\,,
    \label{eq:qL-AB} \\
    q_{L,\Omega}^{C'}(x) &= \frac{d}{dx}\frac{N_c}{4I}\sum_{\substack{m = \mathrm{all},\,n\in\mathrm{occ} \\
    (E_m \neq E_n)}}\mel{n}{\tau^3}{m}\mel{m}{\mathrm{O}^L_n\!(x)}{n}\,,
    \label{eq:qL-C} \\
    q_{L,\eta}(x) &= M_{\mathrm{cl}}\frac{N_c}{I}\!\sum_{\substack{m = \mathrm{all},\,n\in\mathrm{occ} \\
    (E_m \neq E_n)}}\!\frac{\mel{n}{Y^3}{m}\mel{m}{\mathrm{O}^L_n\!(x)}{n}}{E_n - E_m} - \frac{d}{dx}\frac{N_c}{2I}\!\sum_{\substack{m = \mathrm{all},\,n\in\mathrm{occ} \\
    (E_m = E_n)}}\!\mel{n}{Y^3}{m}\mel{m}{\mathrm{O}^L_n\!(x)}{n}\,.
    \label{eq:qL-eta}
\end{align}
\end{widetext}

To identify how the induced singlet field modifies the quark helicity and OAM distributions, we organize each distribution into two contributions, labeled by $\Omega$ and $\eta$.  The $\Omega$ contribution combines the conventional $\{A, B\}'$ and $C'$ terms,
\begin{align}
    &[\Delta u + \Delta d]_\Omega(x) \notag \\
    &\qquad = [\Delta u + \Delta d]_\Omega^{\{A, B\}'}(x) + [\Delta u + \Delta d]_\Omega^{C'}(x)\,,
    \label{eq:pol-omega} \\
    &q_{L, \Omega}(x) = q_{L, \Omega}^{\{A, B\}'}(x) + q_{L, \Omega}^{C'}(x)\,,
    \label{eq:qL-omega}
\end{align}
whereas the $\eta$ contribution is given by Eqs.~\eqref{eq:pol-eta} and \eqref{eq:qL-eta}.  Thus, the former includes both the insertion of $V_\Omega$ in the quark propagator and the expansion of the endpoint rotation matrices, while the latter arises from the insertion of $V_\eta$.

Importantly, the $\Omega$ contribution is not independent of the singlet field.  Although its spectral numerator contains no explicit $\eta(r)$, its normalization involves the total moment of inertia $I = I_{\Omega\Omega} + I_\eta$ in Eq.~\eqref{eq:total-inertia}, rather than $I_{\Omega\Omega}$ alone.  The same total inertia enters the $\eta$ contribution.  The singlet field therefore affects the distributions both through an explicit vertex insertion and through the normalization of the collective rotation.

The first moments of these components are denoted by
\begin{align}
    \Delta\Sigma_a &= \int_{-1}^{1} dx\,[\Delta u+\Delta d]_a(x)\,, \\
    L_{q,a} &= \int_{-1}^{1} dx\,q_{L,a}(x)\,, \qquad a = \Omega, \eta\,.
\end{align}
With $\Delta\Sigma = \Delta\Sigma_\Omega + \Delta\Sigma_\eta$ and $L_q = L_{q, \Omega} + L_{q, \eta}$, the nucleon spin sum rule takes the form
\begin{equation}
    \underbrace{\frac{1}{2}\Delta\Sigma_\Omega + L_{q,\Omega}}_{\Omega\ \mathrm{contribution}} + \underbrace{\frac{1}{2}\Delta\Sigma_\eta + L_{q,\eta}}_{\eta\ \mathrm{contribution}} = \frac{1}{2}\,.
    \label{eq:spinsum}
\end{equation}
As in the conventional CQSM~\cite{Wakamatsu:1990ud, Wakamatsu:1999nj}, the nucleon spin at the model scale is carried entirely by the spin and OAM of quarks and antiquarks, with no explicit gluon contribution.   The $\eta$ components describe the singlet-induced response of these quark observables, not an additional meson contribution to be added to the spin sum.

\section{Numerical calculation}\label{sec:results}

In this section, we evaluate the quark helicity and OAM distributions numerically and examine the effect of the singlet field on the nucleon spin decomposition.

\subsection{Procedure}

To evaluate the quark helicity and OAM distributions, we first determine the pion profile $F(r)$ and the rotation-induced singlet profile $\eta(r)$.  
The pion profile is obtained by imposing stationarity of the classical soliton mass in Eq.~\eqref{eq:Mcl}, with the regularization specified in Sec.~\ref{sec:reg}.  We solve this self-consistency problem iteratively~\cite{Wakamatsu:1998rx, Kubota:1999hx}.  At each iteration, the Hamiltonian in Eq.~\eqref{eq:dirac-spectrum} is diagonalized in each $(K, \Pi)$ sector using the Kahana--Ripka basis~\cite{Kahana:1984be} in a spherical box of radius $D=30/M$.  The resulting eigenstates are used to update $F(r)$, and this procedure is repeated until self-consistency is reached.

In the resulting hedgehog background, we evaluate the source $j_\eta(r)$ and the response kernel $\mathcal{K}(r, r')$ defined in Eqs.~\eqref{eq:eta-source} and \eqref{eq:eta-kernel}.  The singlet profile is then obtained by solving Eq.~\eqref{eq:eta-response-equation}.  For the large-$r$ behavior of both profiles, we replace the numerical solutions by Yukawa-type tails beyond $r=4$~fm.  The decay parameter is set to $200$~MeV for $F(r)$~\cite{Ossmann:2004bp}.  For $\eta(r)$, it is set to the singlet mass $m_0$, except at $m_0=0$, where we use $200$~MeV.

With these profiles and the single-quark eigenstates, we evaluate the helicity distribution $[\Delta u+\Delta d](x)$ from Eqs.~\eqref{eq:pol-AB}--\eqref{eq:pol-eta} and the OAM distribution $q_L(x)$ from Eqs.~\eqref{eq:qL-AB}--\eqref{eq:qL-eta}.  The moment of inertia entering these expressions is the total inertia $I$ of Eq.~\eqref{eq:total-inertia}, including the singlet contribution.  Before evaluating the single-quark matrix elements, we perform the angular average over the spatial direction of the light-cone vector in these expressions.  We use the displayed occupied-state representations for $x\geq0$ and the corresponding unoccupied-state representations for $x<0$.

To obtain smooth distributions from the discrete finite-box spectrum, we smear the delta function $\delta_n$ defined in Eq.~\eqref{eq:delta-n} with a normalized Gaussian of width $\gamma=0.10$~\cite{Diakonov:1997vc, Ossmann:2004bp}.  The first moments $\Delta\Sigma$ and $L_q$, defined in Eqs.~\eqref{eq:DeltaSigma} and \eqref{eq:Lq}, are obtained by integrating the smeared distributions.

\subsection{Parameters and static quantities}

\begin{figure}
\includegraphics[width=\columnwidth]{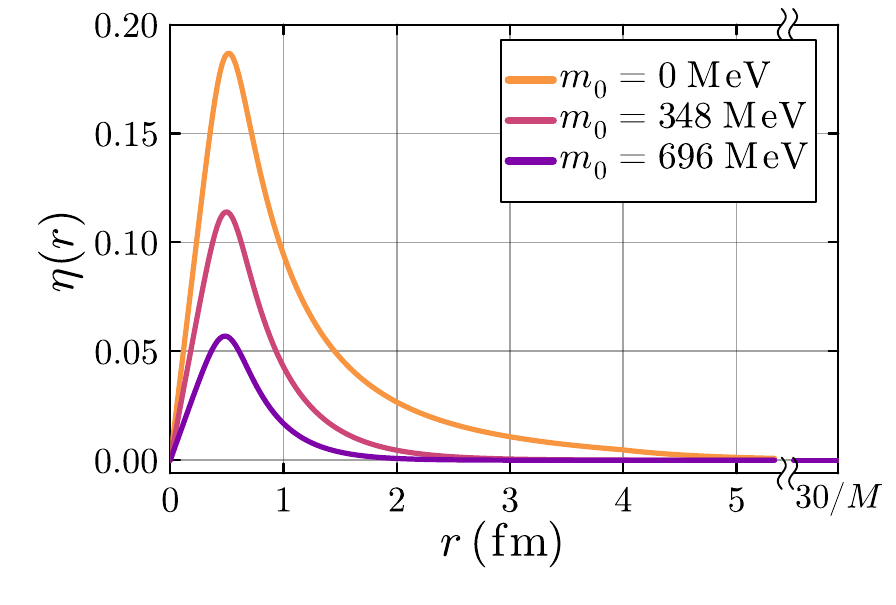}
\caption{The induced profile $\eta(r)$ for the three values of $m_0$, after the tail replacement described in the text.}
\label{fig:eta}
\end{figure}

We work in the chiral limit with $M = 350$~MeV, $f_\pi = 93$~MeV, and $\langle\bar{q}q\rangle = - (286.5~\mathrm{MeV})^3$, which give $\Lambda_1 = 634.05$~MeV and $\Lambda_2 = 1505.53$~MeV.  We set $f_0 = f_\pi$ for simplicity.  For $m_0$ we take the Witten--Veneziano value at $N_f = 2$, $m_0^2 = \tfrac{2}{3}\left(m_\eta^2 + m_{\eta'}^2 - 2m_K^2\right) = (696~\mathrm{MeV})^2$~\cite{Witten:1979vv, Veneziano:1979ec, ParticleDataGroup:2026mpi}, the factor $\tfrac23$ being the ratio of $N_f$ in Eq.~\eqref{eq:m0-chi}.  Since $m_0$ enters only the local diagonal term of Eq.~\eqref{eq:eta-kernel}, the profile $F(r)$, $M_{\mathrm{cl}}$ and $I_{\Omega\Omega}$ do not depend on it, and only $\eta(r)$ and $I_\eta$ change.  Figure~\ref{fig:eta} shows $\eta(r)$ for the three values.  Lowering $m_0$ enhances the peak of $\eta(r)$ and broadens the profile.

The soliton has $M_{\mathrm{cl}}=1062$~MeV and $I_{\Omega\Omega} = 7.69\times10^{-3}$~MeV$^{-1}$.  The $\eta$ field adds $I_\eta = 1.87\times10^{-4}$~MeV$^{-1}$ at $m_0 = 696$~MeV, $4.10\times10^{-4}$~MeV$^{-1}$ at $m_0 = 348$~MeV, and $7.71\times10^{-4}$~MeV$^{-1}$ at $m_0 = 0$.

\subsection{Distributions, first moments and the spin sum rule}

\begin{figure}
\includegraphics[width=\columnwidth]{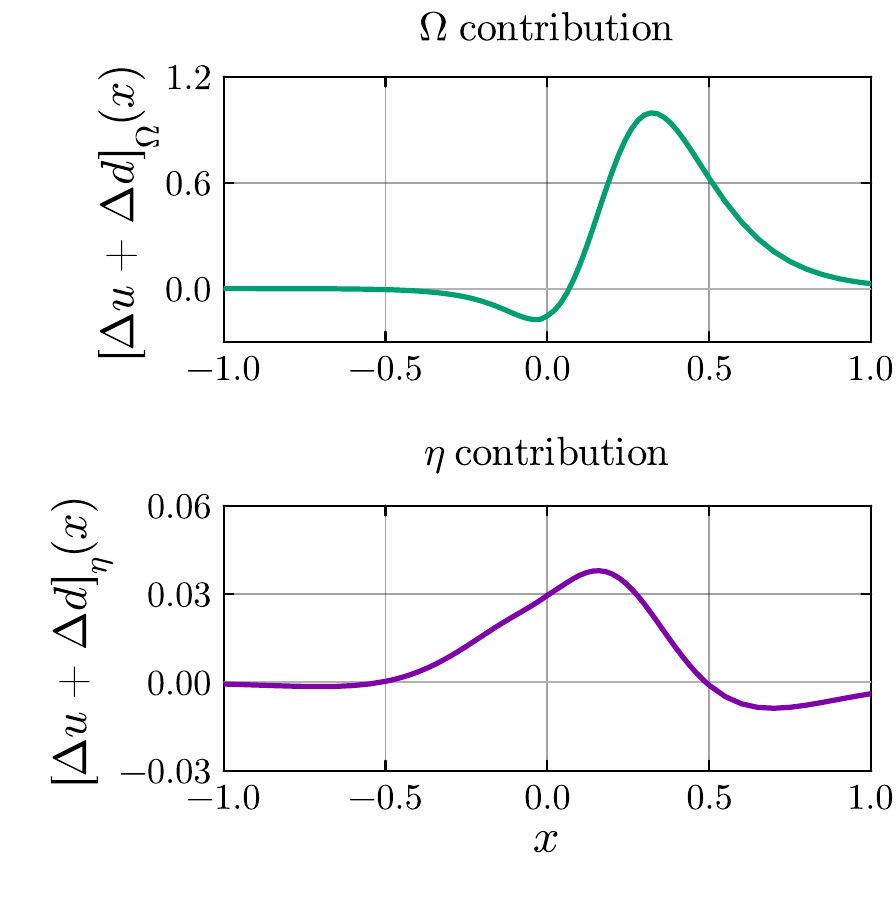}
\caption{$[\Delta u + \Delta d](x)$ from the $\Omega$ contribution (upper panel) and from the $\eta$ contribution (lower panel).  $M = 350$~MeV, $D = 30 / M$, $\gamma = 0.10$ and $m_0 = 696$~MeV.}
\label{fig:dudd}
\end{figure}

\begin{figure}
\includegraphics[width=\columnwidth]{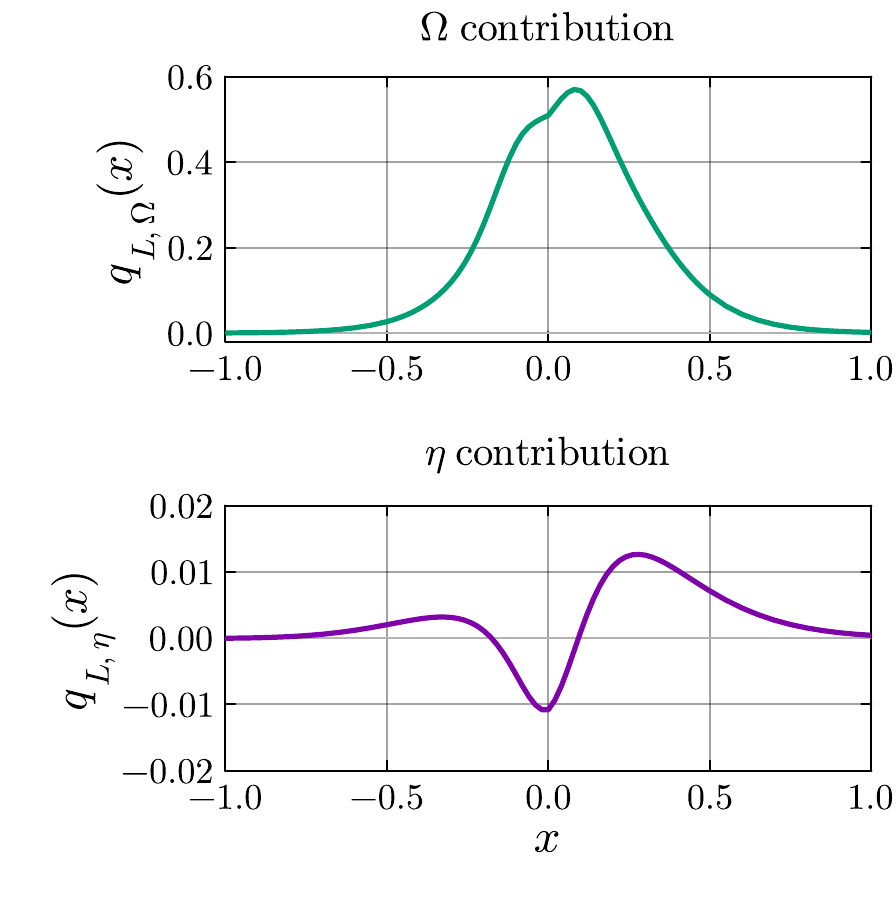}
\caption{$q_L(x)$, in the same decomposition and for the same parameters as Fig.~\ref{fig:dudd}.}
\label{fig:qL}
\end{figure}

Figures~\ref{fig:dudd} and \ref{fig:qL} show the quark helicity and OAM distributions, respectively.  The two panels of each figure use different vertical scales, since the $\eta$ contribution is one to two orders of magnitude smaller.

The first moments and $\frac{1}{2}\Delta\Sigma + L_q$ are collected in Table~\ref{tab:moments}.  With $m_0 = 696$~MeV the $\eta$ contribution is a few percent of $\Delta\Sigma$ and $L_q$.
\begin{table}
\caption{First moments of the two distributions and the spin sum $\frac{1}{2}\Delta\Sigma+L_q$, separated into the $\Omega$ and $\eta$ contributions, with their shares of the total in parentheses.  By Eq.~\eqref{eq:spinsum} the total of $\frac{1}{2}\Delta\Sigma + L_q$ should equal $1/2$.}
\label{tab:moments}
\begin{ruledtabular}
\begin{tabular}{lccc}
$m_0\to\infty$ & $\Omega$ & $\eta$ & total \\
\hline
$\Delta\Sigma$ & $0.373$ & 0 & $0.373$ \\
$L_q$ & $0.315$ & 0 & $0.315$ \\
$\frac{1}{2}\Delta\Sigma + L_q$ & $0.501$ & 0 & $0.501$ \\
\hline
$m_0 = 696$ MeV & $\Omega$ & $\eta$ & total \\
\hline
$\Delta\Sigma$ & $0.364\ (96.0\%)$ & $0.015\ (4.0\%)$ & $0.379$ \\
$L_q$ & $0.307\ (98.5\%)$ & $0.005\ (1.5\%)$ & $0.312$ \\
$\tfrac12\Delta\Sigma+L_q$                  & $0.489\ (97.5\%)$ & $0.012\ (2.5\%)$ & $0.501$ \\
\hline
$m_0 = 348$ MeV & $\Omega$ & $\eta$ & total \\
\hline
$\Delta\Sigma$ & $0.354\ (91.1\%)$ & $0.035\ (8.9\%)$ & $0.389$ \\
$L_q$ & $0.299\ (97.1\%)$ & $0.009\ (2.9\%)$ & $0.307$ \\
$\frac{1}{2}\Delta\Sigma + L_q$ & $0.476\ (94.8\%)$ & $0.026\ (5.2\%)$ & $0.502$ \\
\hline
$m_0 = 0$ MeV & $\Omega$ & $\eta$ & total \\
\hline
$\Delta\Sigma$ & $0.339\ (83.3\%)$ & $0.068\ (16.7\%)$ & $0.407$ \\
$L_q$ & $0.286\ (95.6\%)$ & $0.013\ (4.4\%)$ & $0.299$ \\
$\frac{1}{2}\Delta\Sigma + L_q$ & $0.455\ (90.6\%)$ & $0.047\ (9.4\%)$ & $0.502$ \\
\end{tabular}
\end{ruledtabular}
\end{table}

\begin{figure}
\includegraphics[width=\columnwidth]{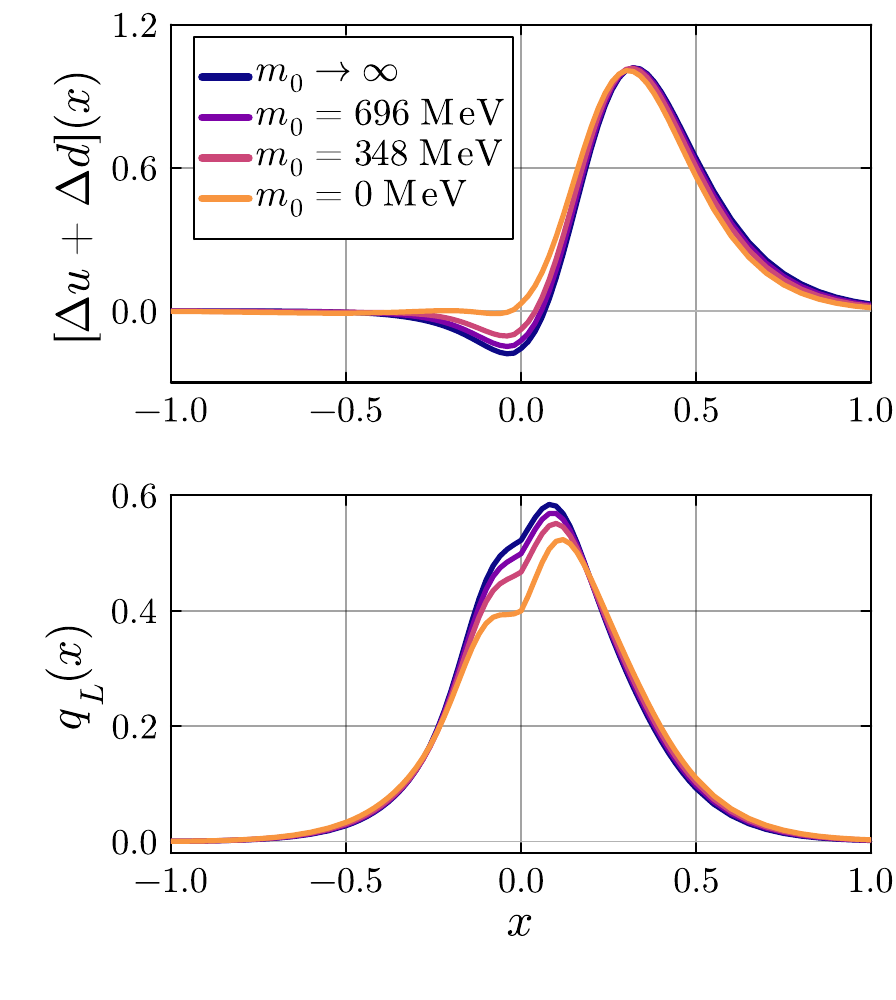}
\caption{$[\Delta u + \Delta d](x)$ (upper panel) and $q_L(x)$ (lower panel), summed over the $\Omega$ and $\eta$ contributions, for $m_0\to\infty$, $696$~MeV, $348$~MeV, and $0$~MeV.  For $m_0\to\infty$, only the $\Omega$ contribution remains, with $I=I_{\Omega\Omega}$.  The other parameters are as in Fig.~\ref{fig:dudd}.}
\label{fig:m0total}
\end{figure}

At $m_0 = 348$~MeV, $I_\eta$ and each $\eta$ contribution are roughly twice their values at $m_0 = 696$~MeV.

\begin{figure}
\includegraphics[width=\columnwidth]{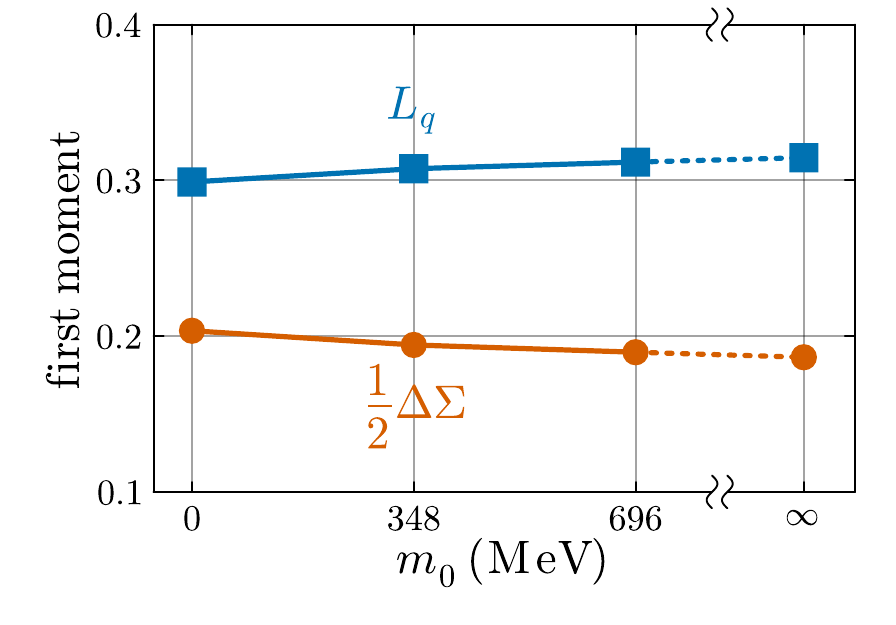}
\caption{$\tfrac12\Delta\Sigma$ (circles) and $L_q$ (squares), summed over the $\Omega$ and $\eta$ contributions, as functions of $m_0$.  The horizontal axis is cut between $m_0=696$~MeV and $m_0\to\infty$.  The solid lines connect the three finite values of $m_0$, and the dotted lines lead to the $m_0\to\infty$ values.  The values are those of Table~\ref{tab:moments}.}
\label{fig:m0moments}
\end{figure}

Figure~\ref{fig:m0total} compares the totals of the two distributions for the four cases of $m_0$ in Table~\ref{tab:moments}, and Fig.~\ref{fig:m0moments} shows how the first moments change with $m_0$.  As $m_0$ decreases from infinity to zero, $\frac{1}{2}\Delta\Sigma$ rises from $0.186$ to $0.203$ and $L_q$ falls from $0.315$ to $0.299$.  The two changes nearly cancel in the sum.

The spin sum rule of Eq.~\eqref{eq:spinsum} is reproduced to better than $1\%$ at all four values of $m_0$.  In the limit $m_0\to\infty$, where $\eta_0$ decouples, $\frac{1}{2}\Delta\Sigma + L_q = 0.501$.

\section{Discussion}\label{sec:discussions}

The results in Table~\ref{tab:moments} and Fig.~\ref{fig:m0moments} show that increasing the anomaly-induced singlet mass reduces the quark helicity and increases the OAM over the parameter values considered.  Since $m_0^2 = 4\chi_{\mathrm{YM}}/f_0^2$, varying $m_0$ at fixed $f_0$ probes the sensitivity of the spin decomposition to anomalous $U(1)_A$ breaking in the effective theory.  The observed dependence can be understood as the suppression, by the anomalous mass, of a rotation-induced singlet response that enhances the quark helicity.

To expose the coupling of the singlet field to quark spin, consider the local field redefinition
\begin{equation}
    \psi = e^{-i\gamma_5\eta_0/(2f_0)}q\,, \qquad\bar\psi = \bar q\,e^{-i\gamma_5\eta_0/(2f_0)}\,.
    \label{eq:discussion-chiral-rotation}
\end{equation}
At the level of the quark bilinear in Eq.~\eqref{eq:cqsm-lagrangian}, this removes the singlet phase from the constituent-quark mass term and replaces it with a derivative coupling,
\begin{equation}
    \Lag_q = \bar q\left(i\Slash{\partial} - MU^{\gamma_5}\right)q + \frac{\partial_\mu\eta_0}{2f_0}J_{5, q}^{\mu}\,,
    \label{eq:discussion-derivative-coupling}
\end{equation}
where
\begin{equation}
    J_{5, q}^{\mu} = \bar q\gamma^\mu\gamma_5q = \bar\psi\gamma^\mu\gamma_5\psi
    \label{eq:discussion-axial-current}
\end{equation}
is the flavor-singlet axial current.

The spatial components of this current are directly related to the quark-spin density, $s_q^i$, i.e.,
\begin{equation}
    J_{5, q}^{i} = 2s_q^i\,, \qquad s_q^i=\frac{1}{2}q^\dagger\gamma^0\gamma^i\gamma_5q\,.
    \label{eq:discussion-spin-density}
\end{equation}
The spatial part of the derivative coupling gives the quark-spin interaction Hamiltonian
\begin{equation}
    H_{\eta s} = - \frac{1}{f_0}\int d^3r\,\bm{\nabla}\eta_0(\bm{r})\cdot\bm{s}_q(\bm{r})\,.
    \label{eq:discussion-spin-hamiltonian}
\end{equation}
Thus, the gradient of the rotation-induced singlet field in Eq.~\eqref{eq:eta-profile} couples directly to the flavor-singlet quark-spin density.  This coupling modifies the quark-spin expectation value.  Since the local axial current is unchanged by Eq.~\eqref{eq:discussion-chiral-rotation}, this is the same spin matrix element that determines $\Delta\Sigma$.

A smaller anomalous mass $m_0$ lowers the energy cost of inducing a nonzero $\eta_0$ field, allowing a larger response to collective rotation.  The induced field modifies the quark-spin matrix element through Eq.~\eqref{eq:discussion-spin-hamiltonian}.  Although this interaction favors alignment of the quark spin with the local gradient of $\eta_0$, such alignment does not necessarily increase the total quark spin along the nucleon polarization direction.  The net change in $\Delta\Sigma$ must therefore be evaluated with the normalization by the total moment of inertia included.

A complementary spatial interpretation follows from the analogy between Gauss's law and the static anomalous Ward identity, $\bm{\nabla}\cdot\bm{J}_5 = 2N_f q_{\mathrm{top}}$ in the chiral limit.  The topological charge density plays the role of the electric charge density, while the spatial axial current plays the role of the electric field.  Since the nucleon forms a dipole of topological charge~\cite{Fukushima:2026sot}, the quark-spin density within the nucleon is analogous to the electric field generated by an electric dipole.  For a localized axial current with a vanishing surface term at infinity, the magnitude of the topological dipole moment is proportional to that of the spatially integrated axial current, and hence to the quark helicity contribution.  The helicity enhancement found at smaller $m_0$ can therefore be interpreted as a strengthening of the topological dipole.

In the present calculation, lowering $m_0$ increases the explicit $\eta$ contribution to $\Delta\Sigma$ more than it reduces the $\Omega$ contribution through the increase in the moment of inertia.  The net effect is therefore an increase in the quark helicity fraction, accompanied by a decrease in the OAM fraction.  Conversely, stronger anomalous $U(1)_A$ breaking suppresses this helicity enhancement by inhibiting the rotation-induced singlet field.

Quantitatively, however, this redistribution is modest.  Increasing $m_0$ from $348$ to $696$~MeV reduces $\Delta\Sigma$ from $0.389$ to $0.379$, corresponding to a relative reduction of approximately $2.6\%$.  At $m_0 = 696$~MeV, the net difference from the pion-only CQSM value, $\Delta\Sigma = 0.373$, is only $0.006$.  A qualitative interpretation of this weak dependence is suggested by the competing roles of the singlet field and the topological susceptibility.  Although lowering $m_0$ enhances and broadens the induced singlet field, it simultaneously reduces $\chi_{\mathrm{YM}} = f_0^2 m_0^2/4$.  Since the topological charge density is proportional to $\chi_{\mathrm{YM}}\eta_0/f_0$, rather than to $\eta_0$ alone, these effects can partly compensate in its dipole moment.  For nonzero $m_0$, this moment is proportional to $\Delta\Sigma$ through the static anomalous Ward identity, providing a possible spatial interpretation of the modest helicity variation. Thus, while the anomalous singlet mass controls the helicity enhancement, the inclusion of the singlet field does not qualitatively alter the spin decomposition of the conventional CQSM.

\section{Conclusions}
\label{sec:conclusions}

We extended the two-flavor CQSM to include the flavor-singlet pseudoscalar field, $\eta_0$, and its anomaly-induced mass, $m_0$, and calculated the quark helicity and OAM distributions to first order in the collective angular velocity.  The induced singlet field modified these distributions through both an explicit vertex insertion and the normalization by the total moment of inertia.  The first moments satisfied the nucleon spin sum rule.  Increasing $m_0$ with the remaining parameters held fixed reduced the quark helicity contribution and increased the OAM.  This redistribution of the nucleon spin remained modest over the parameter values considered and did not qualitatively alter the spin decomposition of the conventional CQSM.

A natural extension of this work is to calculate the flavor-singlet polarized GPDs, $\widetilde H$ and $\widetilde E$, within the same framework.  This would allow the singlet pseudoscalar dynamics to be investigated at finite momentum transfer, with particular attention to the constraints imposed by the anomalous axial Ward identity.

\begin{acknowledgments}
The authors thank Kenji~Fukushima and Christian~Weiss for useful discussions.  This work was partially supported by IIW, WINGS Program, the University of Tokyo (J.M.) and by FoPM, WINGS Program, the University of Tokyo (T.U.).
\end{acknowledgments}

\bibliography{Skyrme}

\end{document}